\documentclass[runningheads,a4paper]{llncs}

\usepackage{amssymb}
\usepackage{graphicx}
\usepackage{amsmath}
\usepackage{packages/multirow}
\usepackage{packages/bbm}
\usepackage{packages/dsfont}

\usepackage[bottom]{packages/footmisc}

\usepackage{url}
\urldef{\mailsa}\path|{binh, kenro.aihara, takasu}@nii.ac.jp|
\newcommand{\keywords}[1]{\par\addvspace\baselineskip
\noindent\keywordname\enspace\ignorespaces#1}

\begin{document}

\mainmatter

\graphicspath{{images/}}
\title{Collaborative Item Embedding Model for Implicit Feedback Data}
\titlerunning{Collaborative Item Embedding Model for Implicit Feedback Data}

\author{ThaiBinh Nguyen\inst{1}%
\and Kenro Aihara\inst{2}\and Atsuhiro Takasu\inst{2}}
\authorrunning{ThaiBinh Nguyen, Kenro Aihara, Atsuhiro Takasu}

\institute{Department of Informatics,\\SOKENDAI (The Graduate University for Advanced Studies),\\Shonan Village, Hayama, Kanagawa 240-0193 Japan\\
\and
National Institute of Informatics,\\2-1-2 Hitotsubashi, Chiyoda, Tokyo 101-8430, Japan\\
\mailsa\\}

\toctitle{Lecture Notes in Computer Science}
\tocauthor{Authors' Instructions}
\maketitle

\begin{abstract}
Collaborative filtering is the most popular approach for recommender systems. One way to perform collaborative filtering is matrix factorization, which characterizes user preferences and item attributes using latent vectors. These latent vectors are good at capturing global features of users and items but are not strong in capturing local relationships between users or between items. In this work, we propose a method to extract the relationships between items and embed them into the latent vectors of the factorization model. This combines two worlds: matrix factorization for collaborative filtering and item embedding, a similar concept to word embedding in language processing. Our experiments on three real-world datasets show that our proposed method outperforms competing methods on top-$n$ recommendation tasks.
\end{abstract}

\section{Introduction}
Modern recommender systems (RSs) are a core component of many online services. An RS analyzes users' behavior and provides them with personalized recommendations for products or services that meet their needs. For example, Amazon recommends products to users based on their shopping histories; an online newspaper recommends articles to users based on what they have read.

Generally, an RS can be classified into two categories: Content-based approach and collaborative filtering-based (CF-based) approach. The content-based approach creates a description for each item and builds a profile for each user's preferences. In other words, the content-based approach recommends items that are similar to items for which the user has expressed interest in the past. In contrast, the CF-based approach relies on the past behavior of each user, without requiring any information about the items that the users have consumed. An advantage of the CF-based approach is that it does not require collection of item contents or analysis. In this work, we focus on the CF-based approach.

Input data for CF-based methods are the user-item interaction matrix, in which each entry is the feedback of a user to an item. The feedback can be explicit (e.g., rating scores/stars, like/dislike) or implicit (e.g., click, view, purchase). Early work mainly focused on explicit feedback data such as SVD++ \cite{koren2008factorization}, timeSVD \cite{journals/cacm/Koren10}, or probabilistic matrix factorization \cite{salakhutdinov2008a}. One advantage of explicit feedback is that it is easy to interpret because it directly expresses the preferences of users for items. However, explicit feedback is not always available and is extremely scarce, as few users provide explicit feedback.

Implicit feedback, in contrast, is generated in abundance while users interact with the system. However, interpreting the implicit feedback is difficult, because it does not directly express users' opinions about items. For example, a user's click on an item does not mean that he or she likes it; rather, the user may click and then find that he or she does not like the item. On the other hand, even though a user does not interact with an item, this does not imply that the user dislikes it; it may be because the user does not know that the item exists.

Hu et al. proposed the weighted matrix factorization (WMF) \cite{hu2008collaborative}, a special case of the matrix factorization technique targeted to implicit datasets. The model maps each user and item into a low-dimensional vector in a shared latent space, which encodes all the information that describes the user's preference or the item's characteristics. Locations of users and items in the space show their relationships. If two items are close together in the space, they are considered to be similar. On the other hand, if a user and an item are close in the space, that user is considered to like that item.

Detecting the relationships between items is crucial to the performance of the RS. We consider two kinds of relationships, the global relationship and a local one. The former indicates the global structure that relates simultaneously to most or all items, and is captured from the overall information encompassed in all user--item interactions. The latter, in contrast, indicates the relationships between a small set of closely related items \cite{koren2008factorization,DBLP_journals_sigkdd_BellK07}. Detecting the local relationship will benefit the RS in recommending correlated items. One example of correlated items in the movie domain is the three volumes of the film series ``Lord of the Rings.'' Usually, a user who watches one of them will watch the others. The detection of local relationships gives the system the ability to capture such correlations and recommend one of these volumes when it knows that the user has watched the others. However, while WMF as well as other MF-based algorithms are strong at capturing the global relationships, they are poor at detecting the local relationships \cite{koren2008factorization,DBLP_journals_sigkdd_BellK07}.

In this work, we propose a model that can capture both global and local relationships between items. The idea is to extract the relationships between items that frequently occur in the context of each other, and embed these relationships into the factorization model of WMF \cite{hu2008collaborative,pan:icdm08}. The ``context'' can be the items in a user's interaction list (i.e., the items that the user interacts with), or the items in a transaction. Two items are assumed to be similar if they often appear in a context with each other, and their representations should be located close to each other in the space. The proposed model identifies such relationships and reflects them into WMF. This was inspired by word-embedding techniques in natural language processing that represent words by vectors that can capture the relationships between each word and its surrounding words \cite{conf/nips/MikolovSCCD13,mikolov2013efficient,levy2014neural,le2014distributed}.

In detail, we build an item--item matrix containing the context information and embed information from this matrix into the factorization model. The embedding is performed by factorizing the user--item matrix and the item--item matrix simultaneously. In the model, the role of the item--item matrix factorization is to adjust the latent vectors of items to reflect item--item relationships.

The rest of this paper is organized as follows. In Sect. 2, we present the background knowledge related to this work. Section 3 presents the details of our idea and how we add item embedding to the original factorization model. In Sect. 4, we explain our empirical study and the experimental results. After reviewing some related work in Sect. 5, we discuss the results of this work and show some directions for future work in Sect. 6.

\section{Preliminary}
\label{preliminary}
\subsection{Weighted Matrix Factorization}

Suppose we have $N$ users and $M$ items. For each user $u$ and item $i$, we denote by $r_{ui}$ the number of times user $u$ has interacted with item $i$. We assume that user $u$ likes item $i$ if he or she has interacted with item $i$ at least once. For user $u$ and item $i$, we define a reference value $p_{ui}$ indicating whether user $u$ likes item $i$ (i.e., $p_{ui}=1$ if $r_{ui}>0$ and $p_{ui}=0$ otherwise), and a confidence level $c_{ui}$ to represent how confident we are about the value of $p_{ui}$. Following \cite{hu2008collaborative}, we define $c_{ui}$ as:

\begin{equation}
	\label{eq:confidence_level}
	c_{ui}=1+\alpha r_{ui},
\end{equation}
where $\alpha$ is a positive number.

Weighted matrix factorization (WMF) \cite{hu2008collaborative,pan:icdm08}, is a factorization model to learn the latent representations of all users and items in the dataset. The objective function of the model is:
\begin{equation}
      \label{wmf_equation}
    \mathcal{L}(X,Y)=\sum_{u,i}c_{ui}(p_{ui}-\mathbf{x}_u^\top\mathbf{y}_i)^2+\lambda\left(\sum_u||\mathbf{x}_u||^2_F+\sum_i||\mathbf{y}_i||^2_F\right),
\end{equation}
where $X\in\mathbb{R}^{d\times N}$ and $Y\in\mathbb{R}^{d\times M}$ are matrices with columns $\mathbf{x}_u$ and $\mathbf{y}_i$ that are the latent vectors of users and items, respectively; $||.||_F$ is the Frobenius norm of a vector. This optimization problem can be efficiently solved using the Alternating Least Square (ALS) method as described in \cite{hu2008collaborative}.

\subsection{Word Embedding}
\label{sub:word_embedding}
Word embedding models \cite{bengio2003neural,conf/nips/MikolovSCCD13,mikolov2013efficient,le2014distributed} have gained success in many natural language processing tasks. Their goal is to find vector representations of words that can capture their relationship with their context words (i.e., the surrounding words in a sentence or paragraph).

Given a corpus and a word $w$, a context word $c$ of $w$ is a word that occurs within a specific-size window around $w$ (context window) in the corpus. Let $\mathcal{D}$ denote the set of all word--context pairs, i.e., $\mathcal{D} = \{(w,c)|w\in V_W,c\in V_C\}$, where $V_W$ and $V_C$ are the set of words and set of context words, respectively. Word embedding models represent a word $w\in V_W$ and a context word $c\in V_C$ by vectors $\mathbf{w}\in\mathbb{R}^d$ and $\mathbf{c}\in \mathbb{R}^d$, respectively, where $d$ is the embedding's dimensionality.

Mikolov et al. proposed an efficient model for learning word vectors \cite{conf/nips/MikolovSCCD13}, which is performed by maximizing the log-likelihood function for every word-context pair $(w, c)\in\mathcal{D}$:
\begin{equation}
\label{eq:log_sgns_equation}
\log \sigma(\mathbf{w}^\top\mathbf{c})+ k\mathbb{E}_{c_N \propto P_D} \sigma(-\mathbf{w}^\top\mathbf{c}_N),
\end{equation}
where $\sigma(.)$ is the sigmoid function: $\sigma(x)=1/(1+\exp(-x))$, $P_D$ is a distribution for sampling false context words (hence, negative sampling) and $k$ is a hyper-parameter specifying the number of negative samples. This model is called Skip-gram negative sampling (SGNS) \cite{conf/nips/MikolovSCCD13}. Based on this model, Mikolov et al. released a well-known open source package named word2vec\footnote{https://code.google.com/archive/p/word2vec/}.

Levy et al. \cite{levy2014neural} showed that the optimal solutions $\mathbf{w}^*, \mathbf{c}^*$ of Eq. (\ref{eq:log_sgns_equation}) satisfy:
\begin{equation}
\mathbf{w}^{*\top}\mathbf{c}^*=\text{PMI}(w,c)-\log k,
\end{equation}
where $\text{PMI}(w, c)$ is the \textit{pointwise mutual information} between word $w$ and context word $c$. The symbol $k$, again, is the number of negative samples.

The PMI \cite{church90} of a word-context pair $(w, c)$ is a measure that quantifies the association between a word $w$ and a context word $c$. It is defined as:
\begin{equation}
      \text{PMI}(w, c)=\log\frac{P(w,c)}{P(w)P(c)},
\end{equation}
where $P(w, c)$ is the probability that $c$ appears in the context of $w$; $P(w)$ and $P(c)$ are the probabilities that word $w$ and context word $c$ appear in the corpus, respectively. Empirically, PMI can be estimated using the actual number of observations in a corpus:

\begin{equation}
      \text{PMI}(w, c)=\log\left(\frac{\#(w, c)|\mathcal{D}|}{\#(w)\#(c)}\right),
\end{equation}
where $\mathcal{|D|}$ is the size of $\mathcal{D}$; $\#(w,c)$ is the number of times the pair $(w,c)$ appears in $\mathcal{D}$; and $\#(w)=\sum_c\#(w,c)$ and $\#(c)=\sum_w\#(w,c)$ are the numbers of times $w$ and $c$ appear in $\mathcal{D}$, respectively.

Levy et al. \cite{levy2014neural} then proposed a word embedding model by factorizing the matrix $S$, which has elements $S_{wc}$ that are defined in Eq. (\ref{eq:sppmimatrix}). This matrix is called the shifted positive pointwise mutual information matrix (SPPMI matrix).
\begin{equation}
	\label{eq:sppmimatrix}
      S_{wc} = \max\{\text{PMI}(w,c)-\log k, 0\}.
\end{equation}

In other words, the SPPMI matrix $S$ is obtained by shifting the PMI matrix by $\log k$ and then replacing all negative values with zeroes (hence, shifted positive pointwise mutual information).

\section{Co-occurrence-based Item Embedding for Collaborative Filtering}
\label{methodology}
\subsection{Co-occurrence-based Item Embedding}
By considering each item as a word, we aim to extract the relationships between items in the same way as word embedding techniques do. Our motivation is that the representation of an item is governed not only by the users who interact with it but also by the other items that appear in its context. In this work, we define ``context'' as the items occurring in the interaction list of a user (i.e., the items that the user interacts with). However, other definitions of context can also be used without any problems. We argue that if items co-occur frequently in the interaction lists of some users, they are similar, and their latent vectors should be close in the latent space.

Inspired by the work of Levy et al. \cite{levy2014neural}, which we present in Sect. \ref{sub:word_embedding}, we construct an SPPMI matrix of items based on co-occurrences and embed it into the factorization model.

\subsubsection{Constructing the SPPMI matrix for items.} We now show how to construct the SPPMI matrix for items according to their co-occurrences.

Let $\mathcal{D}=\{(i, j)|i, j\in I_u, i\neq j, u\in U\}$, where $I_u$ is the set of items with which user $u$ has interacted. We use $\#(i, j)$ to denote the number of times the item pair $(i, j)$ appears in $\mathcal{D}$ and $\#(i)=\sum_j\#(i,j)$ to denote the number of times item $i$ appears in $\mathcal{D}$.

For example, if we have three users $u_1, u_2$, and $u_3$ whose interaction lists are $I_1=\{1, 2, 4\}$, $I_2=\{2, 3\}$, and $I_3=\{1, 2, 3\}$, respectively, we will have:
\begin{itemize}
\item $\mathcal{D}=\{(1,2), (1,4), (2,4), (2,3), (1,2), (1,3), (2,3)\}$
\item $\#(1, 2)=2, \#(1,3)=1, \#(1,4)=1, \#(2,3)=2, \#(2,4)=1$
\item $\#(1)=4, \#(2)=5, \#(3)=3, \#(4)=2$.
\end{itemize}

The item--item matrix $S$ has elements:
\begin{equation}
	\label{eq:sij}
	s_{ij} = \log\left(\frac{\#(i,j)|\mathcal{D}|}{\#(i)\#(j)}\right)-\log k,
\end{equation}
where $\log\left(\frac{\#(i,j)|\mathcal{D}|}{\#(i)\#(j)}\right)$ is the pointwise mutual information of pair $(i, j)$, as mentioned above, and $k$ is a positive integer corresponding to the number of negative samples in the SGNS model \cite{conf/nips/MikolovSCCD13}. In our experiments, we set $k=1$.

Because $S$ defined above is symmetric, instead of factorizing $S$ into two different matrices as in \cite{levy2014neural}, we factorize it into two equivalent matrices. In more detail, we factorize $S$ to the latent vectors of items:

\begin{equation}
\label{eq:sppmi_item_factorization}
S=Y^\top Y
\end{equation}

In this way, $S$ can also be viewed as a similarity matrix between items, where element $s_{ij}$ indicates the similarity between item $i$ and item $j$.

\subsection{Co-occurrence-based Item Embedded Matrix Factorization (CEMF)}
We can now show how to incorporate the co-occurrence information of items into the factorization model. The SPPMI matrix will be factorized to obtain the latent vectors of items. The learned latent factor vectors of items should minimize the objective function:

\begin{equation}
\label{eq:sppmi_factorization}
\sum_{i,j:s_{ij}>0}\left(s_{ij}-\mathbf{y}_i^\top\mathbf{y}_j\right)^2.
\end{equation}

Combining with the original objective function in Eq. (\ref{wmf_equation}), we obtain the overall objective function:

\begin{multline}
	\label{eq:overal_loss}
  \mathcal{L}(X, Y)=\sum_{u,i}c_{ui}\left(p_{ui}-\mathbf{x}_u^\top\mathbf{y}_i\right)^2+\sum_{\substack{i\\j>i\\ s_{i,j}>0}}\left(s_{ij}-\mathbf{y}_i^\top\mathbf{y}_j\right)^2\\
  +\lambda\left(\sum_u||\mathbf{x}_u||^2_F+\sum_i||\mathbf{y}_i||^2_F\right).
\end{multline}

\subsubsection{Learning method.} This function is not convex with respect to $\mathbf{x}_u$ and $\mathbf{y}_i$, but it is convex if we keep one of these fixed. Therefore, it can be solved using the Alternating Least Square method, similar to the method described in \cite{hu2008collaborative}.

For each user $u$, at each iteration, we calculate the partial derivative of $\mathcal{L}$ with respect to $\mathbf{x}_u$ while fixing other entries. By setting this derivative to zero, $\frac{\partial \mathcal{L}}{\partial \mathbf{x}_u}=0$, we obtain the update rule for $\mathbf{x}_u$:

\begin{align}
\label{eq:xu}
\mathbf{x}_u={}&\left(\sum_i c_{ui}\mathbf{y}_i\mathbf{y}^\top_i+\lambda\mathbf{I}_d\right)^{-1}\left(\sum_i c_{ui}\mathbf{y}_ip_{ui}\right).
\end{align}

Similarly, for each item $i$, we calculate the partial derivative of $\mathcal{L}$ with respect to $\mathbf{y}_i$ while fixing other entries, and set the derivative to zero. We obtain the update rule for $\mathbf{y}_i$:
\begin{equation}
\label{eq:yi}
\begin{aligned}
\mathbf{y}_i={}&\left(\sum_u c_{ui}\mathbf{x}_u\mathbf{x}^\top_u+\sum_{j:s_{i,j}>0} \mathbf{y}_j\mathbf{y}^\top_j + \lambda\mathbf{I}_d\right)^{-1}\\
&\left(\sum_u c_{ui}p_{ui}\mathbf{x}_u+\sum_{j:s_{ij}>0}s_{ij}\mathbf{y}_j\right),
\end{aligned}
\end{equation}

where $\mathbf{I}_d\in\mathbb{R}^{d\times d}$ is the identity matrix (i.e., the matrix with ones on the main diagonal and zeros elsewhere).

\subsubsection{Computational complexity.} For user vectors, as analyzed in \cite{hu2008collaborative}, the complexity for updating $N$ users in an iteration is $\mathcal{O}(d^2\mathcal{|R|}+d^3N)$, where $|\mathcal{R}|$ is the number of nonzero entries of the preference matrix $P$. Since $|\mathcal{R}|>>N$, if $d$ is small, this complexity is linear in the size of the input matrix. For item vector updating, we can easily show that the running time for updating $M$ items in an iteration is $\mathcal{O}(d^2(|\mathcal{R}|+M|\mathcal{S}|)+d^3M)$, where $\mathcal{|S|}$ is the number of nonzero entries of matrix $S$. For systems in which the number of items is not very large, this complexity is not a big problem. However, the computations become significantly expensive for systems with very large numbers of items. Improving the computational complexity of updating item vectors will be part of our future work.

\section{Empirical Study}
\label{experiment}
In this section, we study the performance of CEMF. We compare CEMF with two competing methods for implicit feedback data: WMF \cite{hu2008collaborative,pan:icdm08} and CoFactor \cite{confrecsysLiangACB16}. Across three real-world datasets, CEMF outperformed these competing methods for almost all metrics.

\subsection{Datasets, Metrics, Competing Methods, and Parameter Setting}

\subsubsection{Datasets.} We studied datasets from different domains: movies, music, and location, with varying sizes from small to large. The datasets are:
\begin{itemize}
\item \textit{MovieLens-20M (ML-20M)} \cite{journals/tiis/HarperK16}: a dataset of users' movie ratings collected from MovieLens, an online film service. It contains 20 million ratings in the range 1--5 of 27,000 movies by 138,000 users. We binarized the ratings thresholding at 4 or above. The dataset is available at GroupLens\footnote{https://grouplens.org/datasets/movielens/20m/}.
\item \textit{TasteProfile}: a dataset of counts of song plays by users collected by Echo Nest\footnote{http://the.echonest.com/}. After removing songs that were listened to by less than 50 users, and users who listened to less than 20 songs, we binarized play counts and used them as implicit feedback data.
\item \textit{Online Retail Dataset (OnlineRetail)} \cite{Chen2012}: a dataset of online retail transactions provided at the UCI Machine Learning Repository\footnote{https://archive.ics.uci.edu/ml/datasets/Online+Retail}. It contains all the transactions from December 1, 2010 to December 9, 2011 for a UK-based online retailer.
\end{itemize}

For each user, we selected 20\% of interactions as ground truth for testing. The remaining portions from each user were divided in two parts: 90\% for a training set and 10\% for validation. The statistical information of the training set of each dataset is summarized in Table \ref{training_stats}.

\begin{table}[ht]
\begin{center}
\setlength\tabcolsep{6pt}
\renewcommand{\arraystretch}{1.0}
\caption{Statistical information of the datasets after post-preprocessing}
\label{training_stats}
\begin{tabular}{cccc}
\hline\noalign{\smallskip}
& ML-20M & TasteProfile & OnlineRetail\\
\noalign{\smallskip}
\hline
\noalign{\smallskip}
\# of users            & 138,493 & 629,113 & 3,704 \\
\# of items           & 26,308 & 98,486 & 3,643\\
\# of interactions            & 18M & 35.5M & 235K \\
Sparsity  (\%)  & 99.5 & 99.94 & 98.25 \\
Sparsity of SPPMI matrix  (\%)& 75.42 & 76.34 & 66.24\\
\hline
\end{tabular}
\end{center}
\end{table}

\subsubsection{Evaluation metrics.}
The performance of the learned model was assessed by comparing the recommendation list with the ground-truth items of each user. We used Recall$@n$ and Precision@$n$ as the measures for evaluating the performance.

Recall@$n$ and Precision@$n$ are usually used as metrics in information retrieval. The former metric indicates the percentage of relevant items that are recommended to the users, while the latter indicates the percentage of relevant items in the recommendation lists. They are formulated as:
\begin{equation}
    \label{eq:metric}
    \begin{aligned}
    \text{Recall}@n &= \frac{1}{N}\sum_{u=1}^N\frac{|S_u(n) \cap V_u|}{|V_u|}\\
    \text{Precision}@n &= \frac{1}{N}\sum_{u=1}^N\frac{|S_u(n) \cap V_u|}{n}\\
    \end{aligned}
\end{equation}

where $S_u(n)$ is the list of top-$n$ items recommended to user $u$ by the system and $V_u$ is the list of ground-truth items of user $u$.

\subsubsection{Competing methods.}
We compared CEMF with the following competing methods.

\begin{itemize}
\item \textit{CoFactor} \cite{confrecsysLiangACB16}: factorizes user--item and item--item matrices simultaneously as we do, where the item--item co-occurrence matrix is factorized into two matrices.
\item \textit{WMF} \cite{hu2008collaborative}: a weighted matrix factorization matrix for the implicit feedback dataset.
\end{itemize}

\subsubsection{Parameters.}
\begin{itemize}
\item \textit{Number of factors} $d$: we learn the model with the number of factors running from small to large values: $d=$ \{10, 20, 30, 40, 50, 60, 70, 80, 90, 100\}.
\item \textit{Regularization term}: we set the regularization parameter for the Frobenius norm of user and item vectors as $\lambda=0.01$.
\item \textit{Confidence matrix}: we set $c_{ui}=1+\alpha r_{ui}$ $(\alpha >0)$. We changed the value of $\alpha$ and chose the one that gave the best performance.
\end{itemize}

\subsection{Results}
We evaluated CEMF by considering its overall performance and its performance for different groups of users. Results for Precision@$n$ and Recall@$n$ show that our method outperformed the competing methods.

\subsubsection{Overall performance.} Overall prediction performance with respect to Precision and Recall are shown in Table \ref{precision} and Table \ref{recall} respectively. These are the results for $d=30$; larger values of $d$ produce higher accuracy but the differences in performance between the methods do not change much. The results show that CEMF improves the performances for the three datasets over almost all metrics, except for some metrics with $n>20$ for the \textit{TasteProfile}. If we use only small values of $n$, say $n=5$ or $n=10$, CEMF outperforms all competing methods over the three datasets.

\begin{table}[]
\centering
\setlength\tabcolsep{6pt}
\renewcommand{\arraystretch}{1.1}
\caption{Precision@$n$ of WMF, CoFactor, and CEMF over three datasets}
\label{precision}
\begin{tabular}{|l|lccccc|}
\hline
Dataset                       &    Model      & Pre@5 & Pre@10 & Pre@20 & Pre@50 & Pre@100 \\
\hline
\multirow{3}{5em}{ML-20M}       & WMF      &   0.2176    &   0.1818     &    0.1443    &   0.0974   &  0.0677  \\
                              & CoFactor &   0.2249    &   0.1835    &   0.1416     &    0.0926    &   0.0635      \\
                              & CEMF     &   \textbf{0.2369}    &   \textbf{0.1952}     &   \textbf{0.1523}     &  \textbf{0.1007}      &   \textbf{0.0690}      \\
\hline\hline
\multirow{3}{5em}{TasteProfile} & WMF      &   0.1152	& 0.0950	& 0.0755	& \textbf{0.0525} & \textbf{0.0378}\\
                              & CoFactor &    0.1076	& 0.0886	& 0.0701	&  0.0487 & 0.0353\\
                              & CEMF     &    \textbf{0.1181}	& \textbf{0.0966}	& \textbf{0.0760}	& 0.0523 & 0.0373\\
\hline\hline
\multirow{3}{5em}{OnlineRetail} & WMF      &    0.0870 &	0.0713	& 0.0582	& 0.0406 & 0.0294\\
                              & CoFactor &    0.0927	& 0.0728	& 0.0552	& 0.0381 & 0.0273\\
                              & CEMF     &    \textbf{0.0959}	& \textbf{0.0779}	& \textbf{0.0619}	& \textbf{0.0425} & \textbf{0.0302}\\
\hline
\end{tabular}
\end{table}

\begin{table}[ht]
\centering
\setlength\tabcolsep{3pt}
\renewcommand{\arraystretch}{1.1}
\caption{Recall@$n$ of WMF, CoFactor, and CEMF over three datasets}
\label{recall}
\begin{tabular}{|l|lccccc|}
\hline
Dataset                       &    Model      & Recall@5 & Recall@10 & Recall@20 & Recall@50 & Recall@100 \\
\hline
\multirow{3}{6em}{ML-20M}       & WMF      &   0.2366	 & 0.2601	 & 0.3233	& 0.4553	& 0.5788  \\
                              & CoFactor &   0.2420 &	0.2550	& 0.3022 & 0.4101 & 0.5194 \\
                              & CEMF     &   \textbf{0.2563} & 	\textbf{0.2750} &	\textbf{0.3331} & \textbf{0.4605} & \textbf{0.5806} \\
\hline\hline
\multirow{3}{6em}{TasteProfile} & WMF      &   0.11869& 0.1148	& \textbf{0.1377}	& \textbf{0.2129} & 0.\textbf{2960}\\
                              & CoFactor &    0.1106	& 0.1060	& 0.1256	& 0.1947 & 0.2741\\
                              & CEMF     &   \textbf{0.1215}	& \textbf{0.1159}	& 0.1369	& 0.2092 & 0.2891\\
\hline\hline
\multirow{3}{6em}{OnlineRetail} & WMF      &    0.1142 &	0.1463 &	0.2136 & 	0.3428 & 0.4638 \\
                              & CoFactor &    0.1160	& 0.1384	& 0.1891	&  0.3020 & 0.4159 \\
                              & CEMF     &    \textbf{0.1232}	& \textbf{0.1550}	& \textbf{0.2191}	& \textbf{0.3466} & \textbf{0.4676}\\
\hline
\end{tabular}
\end{table}

\subsubsection{Performance for different groups of users.} We divided the users into groups based on the number of items they had interacted with so far, and evaluated the performance for each group. There were three groups in our experiments:
\begin{itemize}
\item \textit{low}: users who had interacted with less than 20 items
\item \textit{medium}: users who had interacted with $20\sim 100$ items
\item \textit{high}: users who had interacted with more than 100 items.
\end{itemize}

\begin{figure}[htp]
\centering
\includegraphics[width=.46\textwidth]{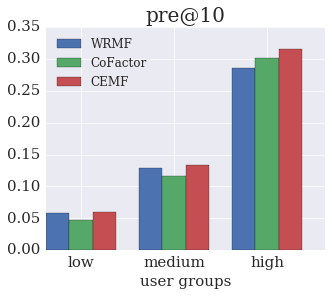}\hfill
\includegraphics[width=.46\textwidth]{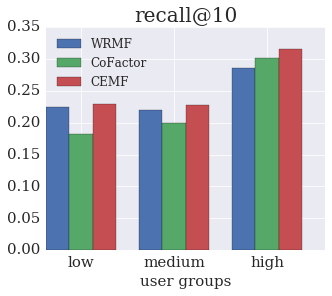}\hfill
\caption{Precision@10 and Recall@10 for different groups of users with the ML-20M dataset}
\label{fig:group_based}

\end{figure}

The Precision@$n$ and Recall@$n$ for these groups are presented in Fig. \ref{fig:group_based}. The results show that CEMF outperforms the competing methods for almost all groups of users. For users with small numbers of interactions, CEMF is slightly better than WMF and much better than CoFactor. For users with many items in their interaction lists, CEMF shows much better performance than WMF and better than CoFactor.

In a system, we usually have users with few interactions and users with many interactions; therefore, using CEMF is more efficient than either WMF or CoFactor.

\section{Related Work}
\label{related_work}
Standard techniques for implicit feedback data include weighted matrix factorization \cite{hu2008collaborative,pan:icdm08}, which is a special case of the matrix factorization technique that is targeted to implicit feedback data, where the weights are defined from the interaction counts, reflecting how confident we are about the preference of a user for an item. Gopalan et al. \cite{journals/corr/GopalanHB13} introduced a Poisson distribution-based factorization model that factorizes the user--item matrix.
The common point of these methods for matrix factorization is that they assume that the user--item interactions are independent; thus, they cannot capture the relationships between strongly related items in the latent representations.

Collective matrix factorization (CMF) \cite{SinghG_kdd08} proposes a framework for factorizing multiple related matrices simultaneously, to exploit information from multiple sources. This approach can incorporate the side information (e.g., genre information of items) into the latent factor model.

In \cite{conf/kdd/ZhengDMZ13}, the authors present a factorization-based method that uses item--item similarity to predict drug--target interactions. While this model uses the item--item similarity from additional sources as side information, we do not require side information in this work. Instead, we exploit the co-occurrence information that is drawn from the interaction matrix.

The CoFactor \cite{confrecsysLiangACB16} model is based on CMF \cite{SinghG_kdd08}. It factorizes the user--item and item--item matrices at the same time in a shared latent space. The main difference between our method and CoFactor is how we factorize the item--item co-occurrence matrix. Instead of representing each item by two latent vectors as in \cite{confrecsysLiangACB16}, where it is difficult to interpret the second one, we represent each item by a single latent vector.

\section{Discussion and Future Work}
\label{conclusion}
We have examined the effect of co-occurrence on the performance of recommendation systems. We proposed a method that combines the power of two worlds: collaborative filtering by MF and item embedding with item context for items in the interaction lists of users. Our goal is a latent factor model that reflects the strong associations of closed related items in their latent vectors. Our proposed method improved the recommendation performance on top-$n$ recommendation for three real-world datasets.

We plan to explore several ways of extending or improving this work. The first direction is to consider different definitions of ``context items''. One approach is to define context items as items that co-occur in the same transactions as the given items. In this way, we can extract relationships between items that frequently appear together in transactions and can recommend the next item given the current one, or recommend a set of items.

The second direction we are planning to pursue is to reduce the computational complexity of the current algorithm. As we mentioned in Sect. \ref{methodology}, the computational complexity for updating item vectors is $\mathcal{O}(d^2(\mathcal{|R|}+M\mathcal{|S|})+d^3M)$, which becomes significantly expensive for systems with large numbers of items. We hope to develop a new algorithm that can improve this complexity. An online learning algorithm, which updates user and item vectors when new data are collected without retraining the model from the beginning, is also in our plan to improve this work.

\subsubsection*{Acknowledgments.} This work was supported by a JSPS Grant-in-Aid for Scientific Research (B) (15H02789).
\bibliographystyle{unsrt}
\bibliography{icwe2017_final}
\end{document}